\documentclass[10pt,prd,aps,twocolumn,nofootinbib,superscriptaddress,notitlepage]{revtex4-1}

\usepackage{amsfonts}
\usepackage{amsmath}
\usepackage{multirow}
\usepackage{array}
\usepackage{graphicx}
\usepackage{bm}
\usepackage{hyperref}

\begin{document}

\title{Deep Learning Calibration of the Quasar UV and X-ray Luminosity Relation for Cosmological Applications}

\author{Jiaze Gao}
\affiliation{Institute of Theoretical Physics, School of Physics, Dalian University of Technology, Dalian 116024, China}

\author{Yun Chen}
\email{chenyun@bao.ac.cn}
\affiliation{National Astronomical Observatories, Chinese Academy of Sciences,
Beijing 100101, China}
\affiliation{College of Astronomy and Space Sciences, University of Chinese Academy of Sciences, Beijing, 100049, China}

\author{Lixin Xu}
\email{lxxu@dlut.edu.cn}
\affiliation{Institute of Theoretical Physics, School of Physics, Dalian University of Technology, Dalian 116024, China}

\author{Jianping Hu}
\affiliation{Ministry of Education Key Laboratory for Nonequilibrium Synthesis and Modulation of Condensed Matter, School of Physics, Xi'an Jiaotong University, Xi'an 710049, China}
\affiliation{Key Laboratory of Modern Astronomy and Astrophysics (Nanjing University), Ministry of Education, Nanjing 210093, China}

\author{Xiaoyue Cao}
\affiliation{Institute for Astrophysics, School of Physics, Zhengzhou University, Zhengzhou, 450001, China}
\date{\today}

\begin{abstract}
Quasars can serve as standardizable candles through an empirical scaling relation between their ultraviolet (UV) and X-ray luminosities. Testing the redshift dependence of this relation is important for their use as high-redshift cosmological probes. In this work, we reconstruct the Hubble diagram of the Pantheon+ sample using the deep learning--based \texttt{LADDER} algorithm and use it as a reference to investigate the quasar scaling relation. Our results are consistent with those from Gaussian process regression and narrow-bin analyses, and provide further statistical support for previously reported concerns about the $z<0.7$ sample. These data should therefore be used with caution in cosmological applications. Within $0.7<z<2.26$, the Bayesian evidence strongly favors the redshift-corrected relation over the standard relation. When the analysis is extended to $0.7<z<3.63$, the narrow-bin results show no significant overall linear trend in the slope $\gamma$. The fitted trends have opposite signs in different subintervals, suggesting possible non-monotonic redshift behavior, although the statistical evidence for the local trends remains limited. The centered intercept shows an overall increasing trend, whose interpretation depends on the adopted distance--redshift relation, while the intrinsic dispersion shows an overall decreasing trend. These results support the presence of redshift-dependent statistical structure in the current quasar sample. Further investigation of its origin and cosmological impact can help refine quasar calibration and error modeling and improve the reliability of quasar cosmology.
\end{abstract}

\maketitle
\begin{center}
	\small
	This is the author-accepted manuscript of
	J.~Gao et al., Phys.\ Rev.\ D \textbf{114}, 123505 (2026).
	The published version is available at
	\href{https://doi.org/10.1103/f5rg-k1y7}
	{https://doi.org/10.1103/f5rg-k1y7}.
	
	\copyright\ 2026 American Physical Society.
\end{center}

\section{\label{sec:introduction}Introduction}

When using standard cosmological probes to constrain the equation of state of dark energy, constraints from cosmic microwave background (CMB) observations alone are limited by parameter degeneracies~\citep{Chen:2018dbv,Planck:2018vyg}. The samples of Type Ia supernovae (SNe Ia) are concentrated at relatively low redshifts~\citep{Riess:2020fzl,Freedman:2021ahq,Riess:2021jrx,Brout:2022vxf}, while baryon acoustic oscillation (BAO) measurements typically provide distance constraints at several effective redshifts~\citep{BOSS:2016wmc,eBOSS:2017cqx,BOSS:2017fdr}. Quasar samples used for cosmological studies contain many objects in the range $0.7\lesssim z\lesssim2.5$ and extend to higher redshifts~\citep{2019NatAs...3..272R,2022MNRAS.515.1795B,2023Ap&SS.368...59F,2026MNRAS.545f1905C}, allowing studies of the distance--redshift relation to reach intervals where SNe Ia are relatively sparse. This makes quasars a powerful probe of dark energy properties and the cosmic expansion history.

Quasars---the brightest persistent sources in the universe---have been observed up to redshifts of $z \sim 7.6$~\citep{Mortlock:2011va,Banados:2017unc,2021ApJ...907L...1W,Risaliti:2026uum}, making them promising high-redshift cosmological probes~\citep{Elvis:2002ja,2022A&A...668A..51L,Wang:2019gaq}. Although quasars do not have constant luminosity, they can be standardized as candles through empirical relations between their luminosity and various observables~\citep{1977ApJ...214..679B,Osmer:1998mb,Watson:2011um,Marziani:2013zra,Wang:2013ha,LaFranca:2014eba,2015ApJ...801....8K,Hu:2020mzd,Wang:2024nsi,2023MNRAS.522.1247K,2026A&A...706A.337L}. In particular, the scaling relation between ultraviolet luminosity ($L_{\mathrm{UV}}$) and X-ray luminosity ($L_{\mathrm{X}}$) in quasars is of high quality~\citep{1979ApJ...234L...9T,1981ApJ...245..357Z,1986ApJ...305...83A,Risaliti:2015zla,Risaliti:2016nqt,Lusso:2020pdb,Lusso:2025bhy}.

By applying appropriate data selection cuts, the intrinsic dispersion of this relation can be reduced to $\delta_{\mathrm{int}} \sim 0.2$ while retaining a large sample size~\citep{Risaliti:2015zla,2016ApJ...819..154L,2017A&A...602A..79L,2019AN....340..267L,Lusso:2020pdb,Lusso:2025bhy}. This provides a solid foundation for studying the properties of the scaling relation and for subsequent cosmological parameter measurements~\citep{Risaliti:2015zla,Risaliti:2016nqt,Lusso:2019akb,Hu:2020mzd,2020MNRAS.496..708L,Khadka:2020vlh,Khadka:2020tlm,Lusso:2020pdb,Li:2023gpp,Raffai:2024xnn,2020MNRAS.492.4456K,2020ApJ...899...71L,2021MNRAS.505.2111L}.

A potential caveat is that if the quasar scaling relation evolves with redshift, the strong degeneracy between cosmological parameters and scaling relation parameters would make it difficult to disentangle cosmological effects from scaling-relation evolution using quasar data alone~\citep{Lusso:2025bhy,Li:2021onq}.

The possible redshift evolution of the quasar scaling relation has been extensively discussed in the literature~\citep{Risaliti:2015zla,Khadka:2020vlh,Khadka:2020tlm,Li:2021onq,Sacchi:2022ofz,Li:2022inq,Wang:2022hko,Risaliti:2023uiy,Li:2024hed,Wang:2024nsi,Wu:2025pmx,2022MNRAS.510.2753K,Lusso:2025bhy}. One class of studies employs cosmology-dependent methods, with the core tension arising from the fact that cosmological parameters inferred from quasar Hubble diagrams are unstable across redshift. Specifically, higher-redshift quasar subsamples tend to yield parameter values that deviate systematically from those derived from lower-redshift data~\citep{2020PhRvD.102l3532Y,Khadka:2020vlh,Khadka:2020tlm,2022MNRAS.510.2753K,2022PhRvD.106d1301O,2025PDU....4901975O}. The other class consists of cosmology-independent methods, including: (1) reconstructing or constraining the luminosity distance using different datasets and analyzing the scaling relation in overlapping redshift intervals, which still reveals redshift-dependent effects~\citep{Li:2021onq,Li:2022inq,Li:2024hed,Wang:2024nsi}; and (2) a narrow-bin method in small luminosity-distance intervals that does not rely on other datasets. With this second method, no systematic redshift evolution is found in the slope $\gamma$ over the redshift range examined~\citep{Risaliti:2015zla,Lusso:2020pdb,Risaliti:2023uiy,Lusso:2025bhy}. Combining this result with other observational evidence, \citet{Lusso:2025bhy} argue that the apparent redshift dependence inferred under an assumed cosmology may instead reflect a mismatch between that model and the actual expansion history. These two cosmology-independent approaches have led to different interpretations, thus posing difficulties for the further development of quasar cosmology.

In this work, we employ \texttt{LADDER}, an open-source neural network based on the Long Short-Term Memory (LSTM) algorithm~\citep{Shah:2024slr}. We train the model on the SNe Ia Pantheon+ dataset and predict distances for the quasar sample at overlapping redshifts in order to investigate the scaling relation in the range $0.01<z<2.26$~\citep{PhysRevD.75.103508,SNLS:2010pgl,2017ApJ...836...56K,Brout:2022vxf,Brout:2021mpj}. We compare the \texttt{LADDER} results with those reported using Gaussian process (GP) regression~\citep{Li:2024hed} and with those from our narrow-bin analysis (binning over small luminosity-distance intervals)~\citep{Lusso:2025bhy}. By comparing the results from these different approaches, we can assess their consistency and gain deeper insight into the redshift dependence of the quasar scaling relation.

The rest of this paper is organized as follows. In Section~\ref{sec:data and ladder}, we describe the quasar and SNe Ia datasets used in this work, along with the models and methods applied. In Section~\ref{sec:result}, we compare \texttt{LADDER} with GP in wide bins, and \texttt{LADDER} with the narrow-bin method in narrow bins, and examine how the parameters of the quasar scaling relation vary with redshift. Section~\ref{sec:summary} summarizes our conclusions.

\section{\label{sec:data and ladder}Data and Methods}

In this section, we describe the quasar and Pantheon+ SNe Ia samples, the \texttt{LADDER} reconstruction method, and the Bayesian framework used to constrain the scaling-relation parameters and compare models.

\subsection{\label{sec:data}Data Sets}
\subsubsection{\label{sec:quasar}Quasar Sample}

The main goal of this work is to study the properties of the scaling relation when using quasars as standardizable candles. We therefore adopt the quasar sample collected and processed by~\citet{Lusso:2020pdb}, which contains approximately 2400 quasars.

This sample was initially compiled from about 19,000 objects drawn from various literature sources and public catalogs~\citep{2016MNRAS.457..110M,2019A&A...632A.109N,Salvestrini:2019thn,2019A&A...630A.118V}. The final sample was obtained after excluding radio-loud sources and broad absorption line quasars, and accounting for dust extinction in the optical and UV bands, host galaxy contamination, gas absorption in X-rays, and Eddington bias related to the X-ray flux limits~\citep{Lusso:2020pdb}.

In this work, we primarily use the sample's spectroscopic redshifts, rest-frame 2500~\AA\ UV flux densities ($F_{\mathrm{UV}}$), and 2~keV X-ray flux densities ($F_{\mathrm{X}}$).

The quasar scaling relation can be expressed as
\begin{equation}\label{eq:Luv-Lx}
    \log {L_{\mathrm{X}}}=\gamma \log {L_{\mathrm{UV}}} + \beta,
\end{equation}
where $\log {L_{\mathrm{X}}}$ and $\log {L_{\mathrm{UV}}}$ denote the base-10 logarithms of the X-ray and ultraviolet luminosities, respectively, $\gamma$ is the slope, and $\beta$ is the intercept.

Using the definition of luminosity distance ($d_{L}$), $L = 4\pi d_{L}^2 F$, the scaling relation can be rewritten in terms of observables as
\begin{equation}\label{eq:Fuv-Fx}
    \log {F_{\mathrm{X}}}=\gamma \log {F_{\mathrm{UV}}} + \hat{\beta},
\end{equation}
where
\begin{equation}\label{eq:beta_hat}
    \hat{\beta}=\beta + (\gamma - 1) \log (4\pi) + 2(\gamma-1)\log d_{L}(z).
\end{equation}

\subsubsection{\label{sec:sn}SNe Ia Sample}

SNe Ia are well-established cosmological probes. In this work, we use the Pantheon+ sample, which includes 1701 light curves of 1550 distinct SNe Ia from 18 different surveys, covering a redshift range of $0.001<z<2.26$~\citep{Brout:2022vxf}. Their apparent magnitudes $m_{B}$ are derived from light-curve fitting using the SALT2 method~\citep{SNLS:2010pgl,Brout:2022vxf,Brout:2021mpj}.

The relation between apparent magnitude and luminosity distance is given by the distance modulus
\begin{equation}\label{eq:mu_obs}
    \mu = m - M,
\end{equation}
where $M$ is the absolute magnitude. The distance modulus is related to the luminosity distance as
\begin{equation}\label{eq:mu_th}
    \mu = 5 \log (d_{L}) + 25,
\end{equation}
with $d_{L}$ in units of Mpc.

\subsection{\label{sec:ladder}\texttt{LADDER} Hubble Diagram}

\texttt{LADDER} (Learning Algorithm for Deep Distance Estimation and Reconstruction)~\citep{Shah:2024slr} is a deep learning algorithm based on LSTM networks. It excels at learning from sequential data and makes use of the full covariance information between data points.

According to~\citet{Shah:2024slr}, \texttt{LADDER} provides a smoother reconstruction and maintains stable performance in sparsely sampled redshift regions. It includes a calibrated gamma-ray burst (GRB) example that demonstrates the algorithm's feasibility for reconstructing the distance ladder.

We use the publicly available \texttt{LADDER} code\footnote{\url{https://github.com/rahulshah1397/LADDER}}, preserving the original training-validation split, to obtain the predicted apparent magnitude--redshift relation from the Pantheon+ sample. To minimize cosmological model dependence in our results, we perform mean-centering of the data using third-order cosmography values ($H_0=73.26$, $q_0=-0.374$, $j_0=-0.023$, $M_B=-19.243$) constrained by Cepheid-calibrated Pantheon+ data~\citep{Riess:2021jrx}, rather than theoretical values from the $\Lambda$CDM model.

We use the released Pantheon+ statistical-plus-systematic covariance matrix and jointly sample the support magnitudes using the corresponding covariance submatrix. The Gaussian Kullback--Leibler (KL) loss is evaluated with natural logarithms, with each prediction matched to its corresponding target. At each requested redshift, we use 100 freshly sampled support realizations and average the predicted means and variances separately, reporting the square root of the averaged variance as the standard deviation. Apart from the adjustments described above, the network architecture and remaining training hyperparameters are unchanged.

\begin{figure}
	\centering
	\includegraphics[scale=0.35]{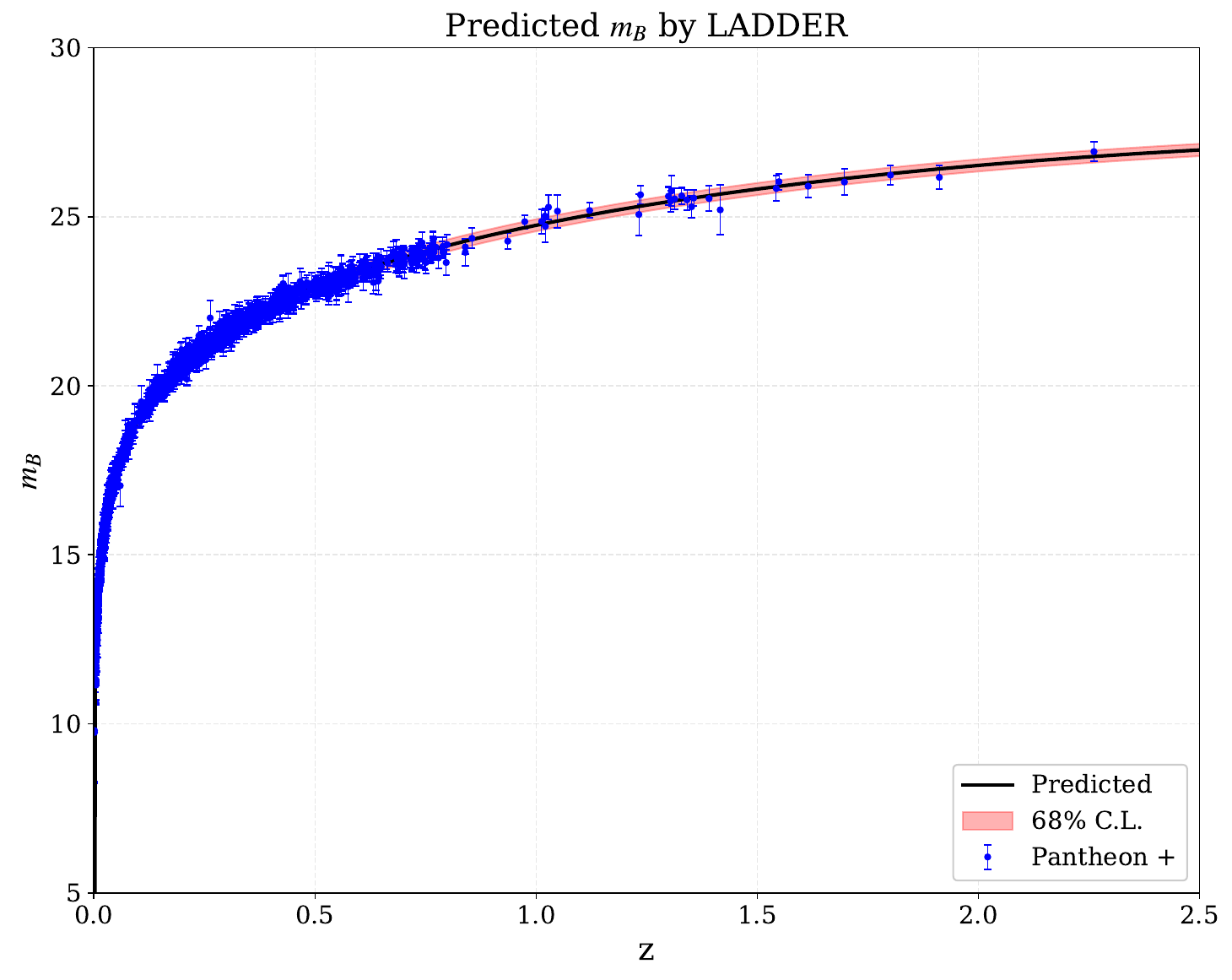}
	\caption{The apparent magnitude $m_B(z)$ of SNe Ia predicted by \texttt{LADDER} (black curve, with the red shaded region indicating the 68\% confidence level) using the Pantheon+ sample, compared against the observed data (blue points). The reconstruction is smooth and closely follows the data across the full redshift range $0.01 < z < 2.26$.}
	\label{fig:mb}
\end{figure}

Figure~\ref{fig:mb} shows the predicted $m_{B}$ together with the Pantheon+ data, illustrating the agreement between predictions and observations, as well as the smoothness of the prediction function at high redshifts. Conservatively, we restrict our analysis to the redshift range covered by SNe Ia, $0.01<z<2.26$. Importantly, the mean-centered loss function is dominated by the Pantheon+ covariance matrix; redshifts beyond the data coverage or in data-scarce regions would be governed by the theoretical model used for mean-centering. This motivates our alternative mean-centering approach.

We evaluate the predicted $\mu(z_i)$ at quasar redshifts $z_i$ using Eq.~(\ref{eq:mu_obs}) and then apply Eqs.~(\ref{eq:mu_th}) and~(\ref{eq:Fuv-Fx}) to analyze the quasar scaling relation. This still requires knowledge of $H_0$ (which is degenerate with $M_B$). We treat $H_0$ as a free parameter with a uniform prior $H_0 \sim \mathcal{U}(50,100)$ (corresponding to $M_B \in (-20.07,-18.57)$).

\subsection{\label{sec:bayesian}Bayesian Inference Framework}

We employ Bayesian inference to estimate the posterior distributions of the quasar scaling relation parameters and to compare models via Bayesian evidence. According to Bayes' theorem, the posterior probability of parameters $\boldsymbol{\theta}$ given data $\mathcal{D}$ and model $\mathcal{M}$ is
\begin{equation}\label{eq:bayesian formula}
    p( \boldsymbol{\theta} \mid \mathcal{D},\mathcal{M}) = \frac{ \mathcal{L}(\mathcal{D} \mid \boldsymbol{\theta},\mathcal{M})  p(\boldsymbol{\theta} \mid \mathcal{M})}{\mathcal{Z} (\mathcal{D} \mid \mathcal{M})},
\end{equation}
where $p(\boldsymbol{\theta} \mid \mathcal{M})$ is the prior distribution, $\mathcal{L}(\mathcal{D} \mid \boldsymbol{\theta},\mathcal{M})$ is the likelihood function, and $\mathcal{Z}(\mathcal{D} \mid \mathcal{M})$ is the Bayesian evidence, obtained by marginalizing over the full parameter space.

In this study, we use the standard likelihood function for standardizable candles:
\begin{small}
\begin{equation}\label{eq:likelihood}
    \mathcal{L} =  { \prod_{i=1}^{N}} \frac {1} {\sqrt{2\pi \sigma_{\mathrm{tot},i}^2}}  \exp \left[ -\frac{(\log F_{\mathrm{X,th},i}-\log F_{\mathrm{X,obs},i})^2}{2{\sigma^2_{\mathrm{tot},i}}}  \right],
\end{equation}
\end{small}
where $\log F_{\mathrm{X,th}}$ is the predicted logarithmic X-ray flux density, and $\sigma_{\mathrm{tot}}$ is the total model uncertainty, obtained either by marginalizing over unknown parameters or by error propagation~\citep{DAgostini:2005mth}.

For the standard (non-evolving) scaling relation, $\log F_{\mathrm{X,th}}$ is computed from Eq.~(\ref{eq:Fuv-Fx}), with the corresponding uncertainty
\begin{equation}
	\sigma^2_{\mathrm{tot}}=\sigma^2_{\log F_\mathrm{X}}+\gamma^2{\sigma^2_{\log F_\mathrm{UV}}}+0.16(\gamma - 1 )^2 \sigma^2_{m_{B}} + \delta^2_{\mathrm{int}},
	\label{eq:classls_error}
\end{equation}
where $\delta_{\mathrm{int}}$ denotes the intrinsic dispersion in $\log F_{\mathrm{X}}$.

This paper also considers a redshift-dependent extension in which the scaling parameters evolve as~\citep{Li:2024hed}
\begin{align}
\gamma \to \gamma_0 + (1+z)\gamma_1, \quad \beta \to \beta_0 + (1+z)\beta_1.
\end{align}
The corresponding theoretical $\log F_\mathrm{X}$ and its uncertainty are computed using these redshift-dependent parameters.

The Bayesian evidence is a prior-weighted average of the likelihood that balances goodness of fit and model complexity, and its natural logarithm, $\ln \mathcal{Z}$, is used for model comparison. For the same data set, the model that provides a better explanation of the data yields a larger $\ln \mathcal{Z}$. The difference in log-evidence (relative Bayesian evidence) determines the relative merit between models~\citep{Kass:1995loi,Trotta:2008qt}.

For the broad-redshift-interval model comparisons, we adopt identical uniform priors for the parameters shared by the two models. The slope parameters $\gamma$ and $\gamma_0$ are assigned $\mathcal{U}(0.01,1)$, and the intercept parameters $\beta$ and $\beta_0$ are assigned $\mathcal{U}(0,15)$, where the subscript $0$ denotes the corresponding coefficients in the redshift-dependent model. The intrinsic dispersion is assigned $\delta_{\rm int}\sim\mathcal{U}(0,1)$. For the additional redshift-dependence coefficients, we adopt $\gamma_1\sim\mathcal{U}(-3,3)$ and $\beta_1\sim\mathcal{U}(-8,8)$.

The subsequent narrow-redshift-bin fits contain relatively few objects in individual bins and therefore yield weaker constraints. We accordingly adopt broader uniform priors, $\beta\sim\mathcal{U}(-10,30)$ and $\hat{\beta}\sim\mathcal{U}(-40,0)$, for the luminosity and flux-space intercepts, respectively, to avoid artificial truncation of the posterior distributions.

In this work, we use the Python open-source package \texttt{PyMultiNest} for nested sampling to obtain posterior parameter distributions and to compute the global Bayesian evidence~\citep{Skilling:2004pqw,10.1111/j.1365-2966.2009.14548.x,Buchner:2014nha}. Visualizations are produced with \texttt{getdist}~\citep{2025JCAP...08..025L}. The posterior distributions of the parameters obtained in this work do not show significant deviations from Gaussianity and are therefore reported in Gaussian form.

\section{\label{sec:result}Results and Discussion}
We use the apparent magnitudes $m_B$ predicted by \texttt{LADDER}---a deep learning algorithm that reconstructs the Hubble diagram directly from the Pantheon+ SNe Ia data without assuming a cosmological model---and treat $M_B$ as a free parameter (which is degenerate with $H_0$). Substituting these quantities into Eq.~(\ref{eq:mu_obs}) and combining with Eqs.~(\ref{eq:mu_th}) and~(\ref{eq:Fuv-Fx}), we infer the posterior distributions of the model parameters using the Bayesian framework described in Section~\ref{sec:bayesian}. This enables a direct comparison of the quasar scaling relation across different redshift intervals. We first perform a wide-bin analysis with the \texttt{LADDER} algorithm in Section~\ref{sec:global fit}. For comparison, we also carry out a narrow redshift binning analysis in Section~\ref{sec:bin}. Finally, in Sec.~\ref{sec:nonlinear}, we further investigate the redshift-dependent structure of the scaling relation and its possible non-monotonic behavior.

\subsection{\label{sec:global fit}Wide-Bin Analysis Using the \texttt{LADDER} Algorithm}

We first consider three different redshift intervals based on sample size and data quality: $0.01<z<2.26$, $0.7<z<2.26$, and $0.7<z<1.7$. Within each interval, we further split the data into two bins according to sample size to enable a comparison between low- and high-redshift subsamples. The constraints on the standard scaling relation obtained from these segmented fits are presented in Table~\ref{table:1}. Results for the standard and redshift-corrected relations fitted directly over the full intervals (without binning) are shown in Table~\ref{table:2}.

\begin{table*}[htbp]
\renewcommand\arraystretch{1.4}
\setlength{\tabcolsep}{10pt}
\footnotesize
\centering
\caption{Constraints on the standard quasar scaling relation (Eq.~\ref{eq:Luv-Lx}) from segmented fits over wide redshift bins, with posterior means and standard deviations. We consider three redshift ranges, each of which is further split into two bins according to sample size. The nuisance parameter $H_0$ is marginalized over with a uniform prior, $H_0\sim\mathcal{U}(50,100)$, and is omitted from subsequent tables for brevity.}
\label{table:1}
\begin{tabular}{cccccccc}
\hline
Subsamples & binning & $\mathcal{N}$
& $\gamma$ & $\beta$ & $\delta_{\mathrm{int}}$
& $H_0$ & $\ln\mathcal{Z}$ \\
\hline
\multirow{2}{*}{$0.01<z<2.26$}
& $0.01<z<1.16$ & $1039$
& $0.654\pm0.015$ & $6.598\pm0.458$ & $0.237\pm0.005$
& $-$ & $-7.492$ \\
& $1.16\leq z<2.26$ & $1026$
& $0.577\pm0.016$ & $8.986\pm0.491$ & $0.225\pm0.005$
& $-$ & $15.784$ \\
\hline\hline
\multirow{2}{*}{$0.7<z<2.26$}
& $0.7<z<1.32$ & $833$
& $0.590\pm0.019$ & $8.537\pm0.579$ & $0.237\pm0.006$
& $-$ & $-13.083$ \\
& $1.32<z<2.26$ & $835$
& $0.566\pm0.017$ & $9.346\pm0.532$ & $0.221\pm0.006$
& $-$ & $21.184$ \\
\hline\hline
\multirow{2}{*}{$0.7<z<1.70$}
& $0.7<z<1.15$ & $625$
& $0.604\pm0.023$ & $8.113\pm0.705$ & $0.239\pm0.007$
& $-$ & $-14.778$ \\
& $1.15<z<1.70$ & $629$
& $0.584\pm0.023$ & $8.754\pm0.694$ & $0.228\pm0.007$
& $-$ & $2.219$ \\
\hline
\end{tabular}
\end{table*}

\begin{figure*}[htbp]
    \centering
    \includegraphics[width=\textwidth]{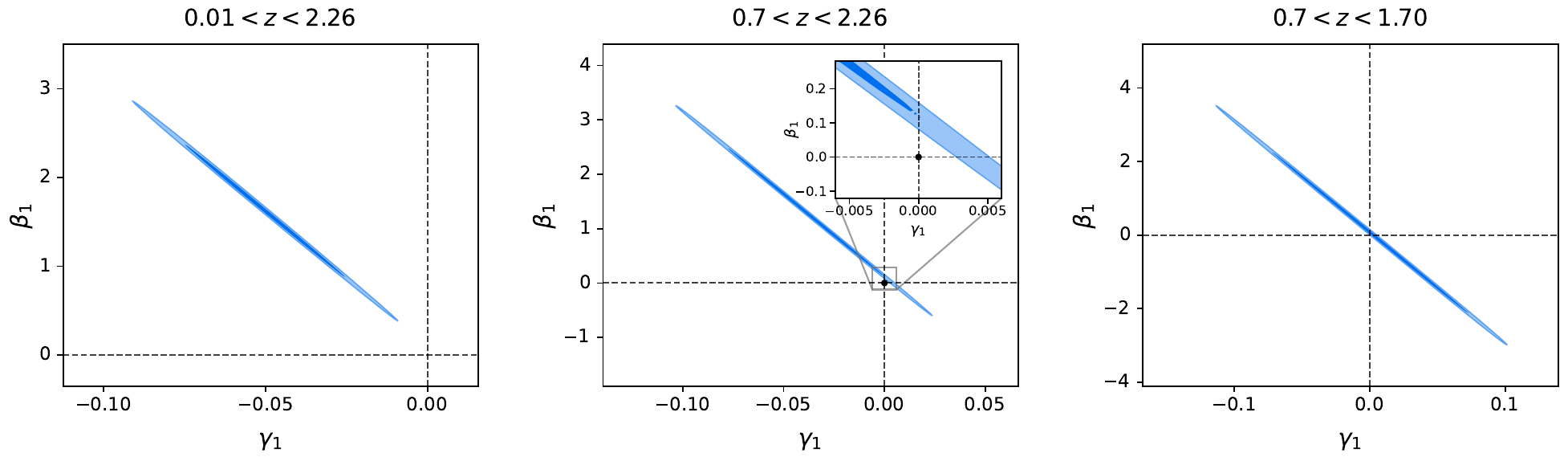}
    \caption{Two-dimensional marginalized posterior distributions
    of the redshift-evolution parameters $\gamma_1$ and $\beta_1$
    for the three quasar samples in Table~\ref{table:2}.
    The dark and light blue regions show the $68\%$ and $95\%$
    credible regions, respectively. The dashed lines mark
    $\gamma_1=0$ and $\beta_1=0$, whose intersection corresponds
    to the classic scaling relation. The inset in the middle
    panel shows an enlarged view around the origin, marked by
    the black point. ChatGPT (6 Astra, OpenAI) assisted with the plotting code; the authors verified the plotted results.}
    \label{fig:gamma1_beta1}
\end{figure*}

\begin{table*}[htbp]
\renewcommand\arraystretch{1.6}
\setlength{\tabcolsep}{7pt}
\footnotesize
\centering
\caption{Fitting results, reported as posterior means and standard deviations, for the wide-bin intervals, comparing the standard scaling relation with a redshift-dependent parameterization
$\gamma=\gamma_0+(1+z)\gamma_1$,
$\beta=\beta_0+(1+z)\beta_1$.
The cutoffs $z=0.7$ and $z=1.7$ are chosen to exclude potentially
contaminated low-redshift quasars and to isolate the region where
previous works reported tension.}
\label{table:2}
\begin{tabular}{cccccccc}
\hline
Subsamples & $L_{\mathrm{UV}}$-$L_{\mathrm{X}}$ relation
& $\gamma$ ($\gamma_0$) & $\beta$ ($\beta_0$)
& $\gamma_1$ & $\beta_1$
& $\delta_{\mathrm{int}}$ & $\ln\mathcal{Z}$ \\
\hline\hline
\multirow{2}{*}{$0.01<z<2.26$}
& classic relation
& $0.647\pm0.009$ & $6.841\pm0.274$
& $\ldots$ & $\ldots$
& $0.233\pm0.004$ & $5.153$ \\
\cline{2-8}
& $z$ - correction
& $0.692\pm0.038$ & $5.243\pm1.129$
& $-0.051\pm0.017$ & $1.644\pm0.501$
& $0.228\pm0.004$ & $28.385$ \\
\hline\hline
\multirow{2}{*}{$0.7<z<2.26$}
& classic relation
& $0.613\pm0.011$ & $7.883\pm0.355$
& $\ldots$ & $\ldots$
& $0.232\pm0.004$ & $4.377$ \\
\cline{2-8}
& $z$ - correction
& $0.652\pm0.064$ & $6.405\pm1.944$
& $-0.040\pm0.026$ & $1.338\pm0.802$
& $0.227\pm0.004$ & $20.782$ \\
\hline\hline
\multirow{2}{*}{$0.7<z<1.70$}
& classic relation
& $0.606\pm0.014$ & $8.063\pm0.433$
& $\ldots$ & $\ldots$
& $0.234\pm0.005$ & $-3.025$ \\
\cline{2-8}
& $z$ - correction
& $0.583\pm0.102$ & $8.594\pm3.079$
& $-0.001\pm0.047$ & $0.097\pm1.417$
& $0.233\pm0.005$ & $-8.104$ \\
\hline
\end{tabular}
\vspace{0.1cm}
\noindent
\parbox{\linewidth}{%
\footnotesize
\raggedright
Subsamples from three distinct redshift intervals were used to constrain both the classic and $z$-corrected scaling relations for quasars.\par
}
\end{table*}

The three redshift intervals are chosen as follows: $0.01<z<2.26$ is the full redshift range adopted for the \texttt{LADDER}-based analysis; $0.7<z<2.26$ excludes the low-redshift range where UV/optical contamination from poor host-galaxy separation may be present~\citep{Lusso:2020pdb,Lusso:2025bhy}; $0.7<z<1.7$ further cuts at $z>1.7$, a region where either scaling relation evolution or cosmological model deviations might occur.

\citet{Li:2024hed} obtained distance measurements via GP regression to study quasar scaling relations using similar binning. Their results are consistent with the corresponding bins in Tables~\ref{table:1} and~\ref{table:2}, demonstrating agreement between the neural network (\texttt{LADDER}) and GP methods. Small differences in $\beta$ arise from the $\beta$--$H_0$ degeneracy and different $H_0$ priors, which do not affect our conclusions.

From Table~\ref{table:1}, we compute the significance of the parameter differences between the low- and high-redshift bins, defined as the difference divided by the joint standard deviation.
\begin{itemize}
\item[(i)] For the pair $0.01<z<1.16$ and $1.16\leq z<2.26$, the discrepancies in $\gamma$, $\beta$, and $\delta_{\mathrm{int}}$ are $3.53\sigma$, $3.56\sigma$, and $1.60\sigma$, respectively.
\item[(ii)] For $0.7<z<1.32$ and $1.32<z<2.26$, they are $0.94\sigma$, $1.03\sigma$, and $1.91\sigma$.
\item[(iii)] For $0.7<z<1.15$ and $1.15<z<1.7$, they are $0.62\sigma$, $0.65\sigma$, and $1.12\sigma$.
\end{itemize}

Thus, after removing the potentially contaminated $z<0.7$ data, the tension in $\{\gamma,\beta\}$ between the low- and high-redshift bins is significantly reduced, while the tension in $\delta_{\mathrm{int}}$ increases. When further excluding $z>1.7$ data, the tensions in all three parameters ($\gamma$, $\beta$, and $\delta_{\mathrm{int}}$) almost completely disappear.

Table~\ref{table:2} corroborates the analysis of Table~\ref{table:1}. For the full $0.01<z<2.26$ interval, the evolution parameters $\{\gamma_1, \beta_1\}$ deviate from zero at approximately $3\sigma$ significance. The relative Bayesian evidence  $\Delta \ln \mathcal{Z} = 23.23$ decisively favors the redshift-dependent parameterization over the standard one. When the low-redshift sources at $z<0.7$ are excluded ($0.7<z<2.26$ interval), $\{\gamma_1, \beta_1\}$ deviate from zero at less than $2\sigma$, and  $\Delta \ln \mathcal{Z} = 16.40$ still strongly favors the redshift-corrected relation but with reduced strength. For the $0.7<z<1.7$ interval, $\{\gamma_1, \beta_1\}$ are consistent with zero within $1\sigma$, and the classic relation is favored ($\Delta \ln \mathcal{Z} = -5.08$).

We further present the two-dimensional marginalized posteriors of $(\gamma_1,\beta_1)$ for the redshift-corrected model in the three redshift intervals listed in Table~\ref{table:2}, as shown in Fig.~\ref{fig:gamma1_beta1}. The two parameters exhibit a strong negative correlation. For $0.7<z<2.26$, although the marginal posterior means of both parameters differ from zero by less than $2\sigma$, the 95\% two-dimensional credible region excludes the origin $(0,0)$ corresponding to the classic relation. This shows that the individual marginal constraints do not fully capture the joint constraints on the two parameters. Their weak marginal departures from zero are therefore not inconsistent with the strong Bayesian evidence favoring the redshift-corrected model. Under the adopted likelihood and priors, these results support the redshift-corrected relation as a better description of the redshift-dependent statistical structure in the current sample.

Therefore, both \texttt{LADDER} and GP reconstructions yield consistent conclusions, indicating that the results are robust against the choice of reconstruction method. After excluding the $z<0.7$ data, no redshift evolution trend is detected in the quasar scaling relation within $0.7<z<1.7$. However, the Bayesian evidence still strongly supports the redshift-corrected relation in the $0.7<z<2.26$ interval ($\Delta \ln \mathcal{Z} = 16.40$). This difference indicates that the inferred model preference depends on the redshift range considered, motivating a closer examination of the redshift-dependent structure through narrow-bin analyses.

\subsection{\label{sec:bin}Narrow Redshift Binning Analysis}

We also compare \texttt{LADDER} with another cosmology-independent method that does not rely on external data, referred to as the ``narrow-bin method.'' Following \citet{Lusso:2025bhy}, we divide the quasar sample into narrow redshift bins so that $\log d_L$ can be treated as constant within each bin. The bin width is chosen to satisfy two conditions: (i) each redshift interval must contain sufficient statistics---here, we require at least 20 objects per bin; and (ii) the range of $\log d_L$ within each bin must be smaller than the dispersion in the $F_\mathrm{X}$--$F_\mathrm{UV}$ relation. Specifically, we impose $\delta_{\log d_L} \ll \delta_{\mathrm{int}}$, for which we adopt the criterion $\delta_{\log d_L} < 0.07$. Under this condition, $\log d_L$ can be treated as approximately constant within each bin, allowing a direct fit of a constant flux intercept $\hat{\beta}$ per bin (Eq.~\ref{eq:beta_hat})~\citep{Risaliti:2015zla,Lusso:2020pdb,Risaliti:2023uiy,Lusso:2025bhy}.

\begin{table*}[htbp]
\renewcommand\arraystretch{1.2}
\setlength{\tabcolsep}{5pt}
\footnotesize
\centering
\caption{Narrow-bin fitting results from separate runs of the original and centered parameterizations, reported as posterior means and standard deviations. Bins satisfy $\delta_{\log d_L}<0.07$. Original reports the luminosity intercept $\beta$ for \texttt{LADDER} and the directly fitted flux intercept $\hat\beta$ for the narrow-bin method. Centered reports the luminosity normalization $\beta_c$ for both methods at the fixed full-sample mean $C=\langle\log L_{\mathrm{UV}}\rangle\simeq30.428$. A reference flat $\Lambda$CDM cosmology with $H_0=70\,\mathrm{km\,s^{-1}\,Mpc^{-1}}$ and $\Omega_{\mathrm{m0}}=0.3$ is used to calculate $C$ and $\delta_{\log d_L}$ and to define the luminosity normalization $\beta_c$ in the centered narrow-bin fits. \texttt{LADDER} results are reported only for bins within its reconstruction range.}
\label{table:3}
\begin{tabular}{ccclcccccc}
\hline
& \multirow{2}{*}{Bin $[z_{\mathrm{min}},z_{\mathrm{max}})$}
& \multirow{2}{*}{$\mathcal{N}$}
& \multirow{2}{*}{Method}
& \multicolumn{2}{c}{$\gamma$}
& \multicolumn{2}{c}{Intercept}
& \multicolumn{2}{c}{$\delta_{\mathrm{int}}$} \\
\cline{5-10}
& & & & Original & Centered & Original & Centered & Original & Centered \\
\hline
& \multirow{2}{*}{$[0.4250,0.4710)$}
& \multirow{2}{*}{$35$} & \texttt{LADDER}
& $0.73\pm0.10$ & $0.72\pm0.11$
& $4.39\pm3.04$ & $26.57\pm0.12$
& $0.24\pm0.03$ & $0.24\pm0.03$ \\
& & & narrow-bin
& $0.73\pm0.11$ & $0.74\pm0.11$
& $-10.86\pm3.05$ & $26.59\pm0.11$
& $0.24\pm0.03$ & $0.24\pm0.03$ \\
\hline
& \multirow{2}{*}{$[0.4710,0.5208)$}
& \multirow{2}{*}{$37$} & \texttt{LADDER}
& $0.60\pm0.10$ & $0.61\pm0.09$
& $8.08\pm2.84$ & $26.42\pm0.11$
& $0.24\pm0.03$ & $0.24\pm0.03$ \\
& & & narrow-bin
& $0.59\pm0.10$ & $0.59\pm0.10$
& $-14.79\pm2.62$ & $26.43\pm0.09$
& $0.24\pm0.03$ & $0.24\pm0.03$ \\
\hline
& \multirow{2}{*}{$[0.5208,0.5745)$}
& \multirow{2}{*}{$57$} & \texttt{LADDER}
& $0.73\pm0.09$ & $0.73\pm0.09$
& $4.19\pm2.52$ & $26.52\pm0.09$
& $0.24\pm0.02$ & $0.24\pm0.02$ \\
& & & narrow-bin
& $0.73\pm0.09$ & $0.73\pm0.09$
& $-11.07\pm2.45$ & $26.54\pm0.08$
& $0.24\pm0.02$ & $0.24\pm0.02$ \\
\hline
& \multirow{2}{*}{$[0.5745,0.6330)$}
& \multirow{2}{*}{$67$} & \texttt{LADDER}
& $0.67\pm0.06$ & $0.67\pm0.06$
& $5.98\pm1.82$ & $26.50\pm0.08$
& $0.23\pm0.02$ & $0.24\pm0.02$ \\
& & & narrow-bin
& $0.67\pm0.06$ & $0.67\pm0.06$
& $-12.75\pm1.74$ & $26.51\pm0.05$
& $0.23\pm0.02$ & $0.23\pm0.02$ \\
\hline
& \multirow{2}{*}{$[0.6330,0.6970)$}
& \multirow{2}{*}{$78$} & \texttt{LADDER}
& $0.68\pm0.07$ & $0.68\pm0.07$
& $5.84\pm2.15$ & $26.51\pm0.08$
& $0.23\pm0.02$ & $0.23\pm0.02$ \\
& & & narrow-bin
& $0.68\pm0.07$ & $0.68\pm0.07$
& $-12.41\pm1.96$ & $26.53\pm0.05$
& $0.23\pm0.02$ & $0.23\pm0.02$ \\
\hline
& \multirow{2}{*}{$[0.6970,0.7660)$}
& \multirow{2}{*}{$96$} & \texttt{LADDER}
& $0.61\pm0.07$ & $0.61\pm0.07$
& $8.08\pm2.14$ & $26.51\pm0.08$
& $0.26\pm0.02$ & $0.27\pm0.02$ \\
& & & narrow-bin
& $0.62\pm0.07$ & $0.62\pm0.07$
& $-14.21\pm1.97$ & $26.54\pm0.04$
& $0.27\pm0.02$ & $0.27\pm0.02$ \\
\hline
& \multirow{2}{*}{$[0.7660,0.8420)$}
& \multirow{2}{*}{$111$} & \texttt{LADDER}
& $0.58\pm0.05$ & $0.58\pm0.06$
& $8.76\pm1.62$ & $26.44\pm0.08$
& $0.24\pm0.02$ & $0.24\pm0.02$ \\
& & & narrow-bin
& $0.58\pm0.06$ & $0.58\pm0.06$
& $-15.45\pm1.59$ & $26.46\pm0.03$
& $0.24\pm0.02$ & $0.24\pm0.02$ \\
\hline
& \multirow{2}{*}{$[0.8420,0.9260)$}
& \multirow{2}{*}{$115$} & \texttt{LADDER}
& $0.57\pm0.05$ & $0.57\pm0.05$
& $9.06\pm1.44$ & $26.46\pm0.08$
& $0.21\pm0.01$ & $0.21\pm0.01$ \\
& & & narrow-bin
& $0.57\pm0.05$ & $0.57\pm0.05$
& $-15.69\pm1.37$ & $26.48\pm0.03$
& $0.21\pm0.01$ & $0.21\pm0.01$ \\
\hline
& \multirow{2}{*}{$[0.9260,1.0064)$}
& \multirow{2}{*}{$120$} & \texttt{LADDER}
& $0.55\pm0.05$ & $0.55\pm0.05$
& $9.84\pm1.63$ & $26.51\pm0.08$
& $0.25\pm0.02$ & $0.25\pm0.02$ \\
& & & narrow-bin
& $0.55\pm0.05$ & $0.55\pm0.06$
& $-16.25\pm1.48$ & $26.53\pm0.03$
& $0.25\pm0.02$ & $0.25\pm0.02$ \\
\hline
& \multirow{2}{*}{$[1.0064,1.1013)$}
& \multirow{2}{*}{$119$} & \texttt{LADDER}
& $0.72\pm0.07$ & $0.72\pm0.07$
& $4.68\pm2.05$ & $26.55\pm0.05$
& $0.27\pm0.02$ & $0.27\pm0.02$ \\
& & & narrow-bin
& $0.72\pm0.07$ & $0.73\pm0.07$
& $-11.49\pm1.88$ & $26.56\pm0.03$
& $0.27\pm0.02$ & $0.27\pm0.02$ \\
\hline
& \multirow{2}{*}{$[1.1013,1.1816)$}
& \multirow{2}{*}{$120$} & \texttt{LADDER}
& $0.55\pm0.05$ & $0.55\pm0.05$
& $9.71\pm1.58$ & $26.49\pm0.08$
& $0.22\pm0.02$ & $0.22\pm0.02$ \\
& & & narrow-bin
& $0.55\pm0.05$ & $0.55\pm0.05$
& $-16.36\pm1.48$ & $26.52\pm0.02$
& $0.22\pm0.02$ & $0.22\pm0.02$ \\
\hline
& \multirow{2}{*}{$[1.1816,1.2860)$}
& \multirow{2}{*}{$120$} & \texttt{LADDER}
& $0.53\pm0.06$ & $0.54\pm0.06$
& $10.28\pm1.80$ & $26.47\pm0.08$
& $0.24\pm0.02$ & $0.24\pm0.02$ \\
& & & narrow-bin
& $0.53\pm0.06$ & $0.52\pm0.06$
& $-16.98\pm1.66$ & $26.50\pm0.02$
& $0.24\pm0.02$ & $0.24\pm0.02$ \\
\hline
& \multirow{2}{*}{$[1.2860,1.3951)$}
& \multirow{2}{*}{$120$} & \texttt{LADDER}
& $0.53\pm0.05$ & $0.53\pm0.05$
& $10.36\pm1.49$ & $26.50\pm0.08$
& $0.22\pm0.02$ & $0.22\pm0.02$ \\
& & & narrow-bin
& $0.53\pm0.05$ & $0.53\pm0.05$
& $-16.84\pm1.39$ & $26.52\pm0.02$
& $0.23\pm0.02$ & $0.23\pm0.02$ \\
\hline
& \multirow{2}{*}{$[1.3951,1.5096)$}
& \multirow{2}{*}{$120$} & \texttt{LADDER}
& $0.65\pm0.06$ & $0.64\pm0.06$
& $6.87\pm1.70$ & $26.52\pm0.07$
& $0.24\pm0.02$ & $0.24\pm0.02$ \\
& & & narrow-bin
& $0.65\pm0.06$ & $0.65\pm0.06$
& $-13.57\pm1.61$ & $26.54\pm0.02$
& $0.24\pm0.02$ & $0.24\pm0.02$ \\
\hline
& \multirow{2}{*}{$[1.5096,1.6000)$}
& \multirow{2}{*}{$120$} & \texttt{LADDER}
& $0.63\pm0.05$ & $0.62\pm0.05$
& $7.49\pm1.43$ & $26.51\pm0.07$
& $0.20\pm0.01$ & $0.20\pm0.01$ \\
& & & narrow-bin
& $0.63\pm0.05$ & $0.62\pm0.05$
& $-14.30\pm1.34$ & $26.54\pm0.02$
& $0.20\pm0.01$ & $0.20\pm0.01$ \\
\hline
& \multirow{2}{*}{$[1.6000,1.7253)$}
& \multirow{2}{*}{$120$} & \texttt{LADDER}
& $0.52\pm0.05$ & $0.52\pm0.05$
& $10.73\pm1.62$ & $26.57\pm0.08$
& $0.23\pm0.02$ & $0.23\pm0.02$ \\
& & & narrow-bin
& $0.52\pm0.05$ & $0.52\pm0.05$
& $-17.23\pm1.47$ & $26.60\pm0.03$
& $0.23\pm0.02$ & $0.23\pm0.02$ \\
\hline
& \multirow{2}{*}{$[1.7253,1.8700)$}
& \multirow{2}{*}{$120$} & \texttt{LADDER}
& $0.55\pm0.05$ & $0.55\pm0.05$
& $9.91\pm1.64$ & $26.62\pm0.08$
& $0.26\pm0.02$ & $0.26\pm0.02$ \\
& & & narrow-bin
& $0.55\pm0.05$ & $0.55\pm0.05$
& $-16.39\pm1.45$ & $26.65\pm0.03$
& $0.26\pm0.02$ & $0.26\pm0.02$ \\
\hline
& \multirow{2}{*}{$[1.8700,1.9977)$}
& \multirow{2}{*}{$120$} & \texttt{LADDER}
& $0.47\pm0.04$ & $0.47\pm0.04$
& $12.27\pm1.11$ & $26.60\pm0.09$
& $0.17\pm0.01$ & $0.17\pm0.01$ \\
& & & narrow-bin
& $0.47\pm0.04$ & $0.47\pm0.04$
& $-18.65\pm0.97$ & $26.63\pm0.02$
& $0.17\pm0.01$ & $0.17\pm0.01$ \\
\hline
& \multirow{2}{*}{$[1.9977,2.1876)$}
& \multirow{2}{*}{$119$} & \texttt{LADDER}
& $0.53\pm0.04$ & $0.53\pm0.04$
& $10.57\pm1.23$ & $26.62\pm0.08$
& $0.20\pm0.02$ & $0.20\pm0.01$ \\
& & & narrow-bin
& $0.53\pm0.04$ & $0.53\pm0.04$
& $-17.10\pm1.10$ & $26.67\pm0.03$
& $0.20\pm0.02$ & $0.20\pm0.02$ \\
\hline
& \multirow{2}{*}{$[2.1876,2.3930)$}
& \multirow{2}{*}{$119$} & \texttt{LADDER}
& $\ldots$ & $\ldots$
& $\ldots$ & $\ldots$
& $\ldots$ & $\ldots$ \\
& & & narrow-bin
& $0.55\pm0.04$ & $0.55\pm0.04$
& $-16.46\pm1.24$ & $26.65\pm0.03$
& $0.19\pm0.02$ & $0.19\pm0.02$ \\
\hline
& \multirow{2}{*}{$[2.3930,2.6330)$}
& \multirow{2}{*}{$94$} & \texttt{LADDER}
& $\ldots$ & $\ldots$
& $\ldots$ & $\ldots$
& $\ldots$ & $\ldots$ \\
& & & narrow-bin
& $0.51\pm0.05$ & $0.51\pm0.05$
& $-17.62\pm1.30$ & $26.70\pm0.04$
& $0.15\pm0.02$ & $0.15\pm0.02$ \\
\hline
& \multirow{2}{*}{$[2.6330,2.9030)$}
& \multirow{2}{*}{$64$} & \texttt{LADDER}
& $\ldots$ & $\ldots$
& $\ldots$ & $\ldots$
& $\ldots$ & $\ldots$ \\
& & & narrow-bin
& $0.63\pm0.05$ & $0.63\pm0.05$
& $-14.10\pm1.38$ & $26.71\pm0.05$
& $0.16\pm0.02$ & $0.16\pm0.02$ \\
\hline
& \multirow{2}{*}{$[2.9030,3.1970)$}
& \multirow{2}{*}{$48$} & \texttt{LADDER}
& $\ldots$ & $\ldots$
& $\ldots$ & $\ldots$
& $\ldots$ & $\ldots$ \\
& & & narrow-bin
& $0.49\pm0.06$ & $0.49\pm0.06$
& $-18.15\pm1.59$ & $26.77\pm0.07$
& $0.19\pm0.03$ & $0.19\pm0.03$ \\
\hline
& \multirow{2}{*}{$[3.1970,3.6320)$}
& \multirow{2}{*}{$20$} & \texttt{LADDER}
& $\ldots$ & $\ldots$
& $\ldots$ & $\ldots$
& $\ldots$ & $\ldots$ \\
& & & narrow-bin
& $0.63\pm0.11$ & $0.63\pm0.11$
& $-14.34\pm2.96$ & $26.68\pm0.15$
& $0.23\pm0.05$ & $0.23\pm0.05$ \\
\hline
\end{tabular}
\end{table*}

Comparing the narrow-bin method to \texttt{LADDER}: the narrow-bin method assumes no abrupt cosmological transitions within small redshift intervals, but introduces unquantified systematic errors by approximating $\log d_L$ as constant. In contrast, \texttt{LADDER} (and GP) rely on SNe Ia data, which become sparse at high redshifts, but offer considerable flexibility within their reconstruction range and do not require a large number of data points per bin. The key advantage of the narrow-bin method is its ability to study the scaling relation without needing high-redshift cosmological data---a capability that SNe Ia-dependent approaches lack.

We apply this narrow-bin method to quasars at $z<3.63$, constructing bins that satisfy $\delta_{\log d_L}<0.07$ while balancing the number of bins and the sample size per bin. We independently fit the scaling relation in each bin using the narrow-bin method and also perform \texttt{LADDER} fits for bins within the SNe Ia reconstruction range. The values of $\delta_{\log d_L}$ and the luminosity intercept $\beta$ for the narrow-bin method are calculated assuming a reference flat $\Lambda$CDM cosmology with $H_0=70\,\mathrm{km\,s^{-1}\,Mpc^{-1}}$ and $\Omega_{\mathrm m0}=0.3$.

In addition, to reduce the posterior correlation between the slope and intercept, we adopt the centered parameterization
\begin{equation}
\log L_{\mathrm X}
=\gamma\bigl(\log L_{\mathrm{UV}}-C\bigr)+\beta_c,
\qquad
\beta_c=\beta+C\gamma.
\end{equation}
Here, $C\simeq30.428$ is the mean logarithmic UV luminosity of the full sample, calculated using the reference distance relation specified above and held fixed in all redshift bins. For \texttt{LADDER}, the reference cosmology is used only to determine $C$, while the fits still use the reconstructed distances. The centered fits preserve the original joint prior through the corresponding variable transformation.

\begin{figure*}[htbp]
    \centering
    \includegraphics[scale=0.6]{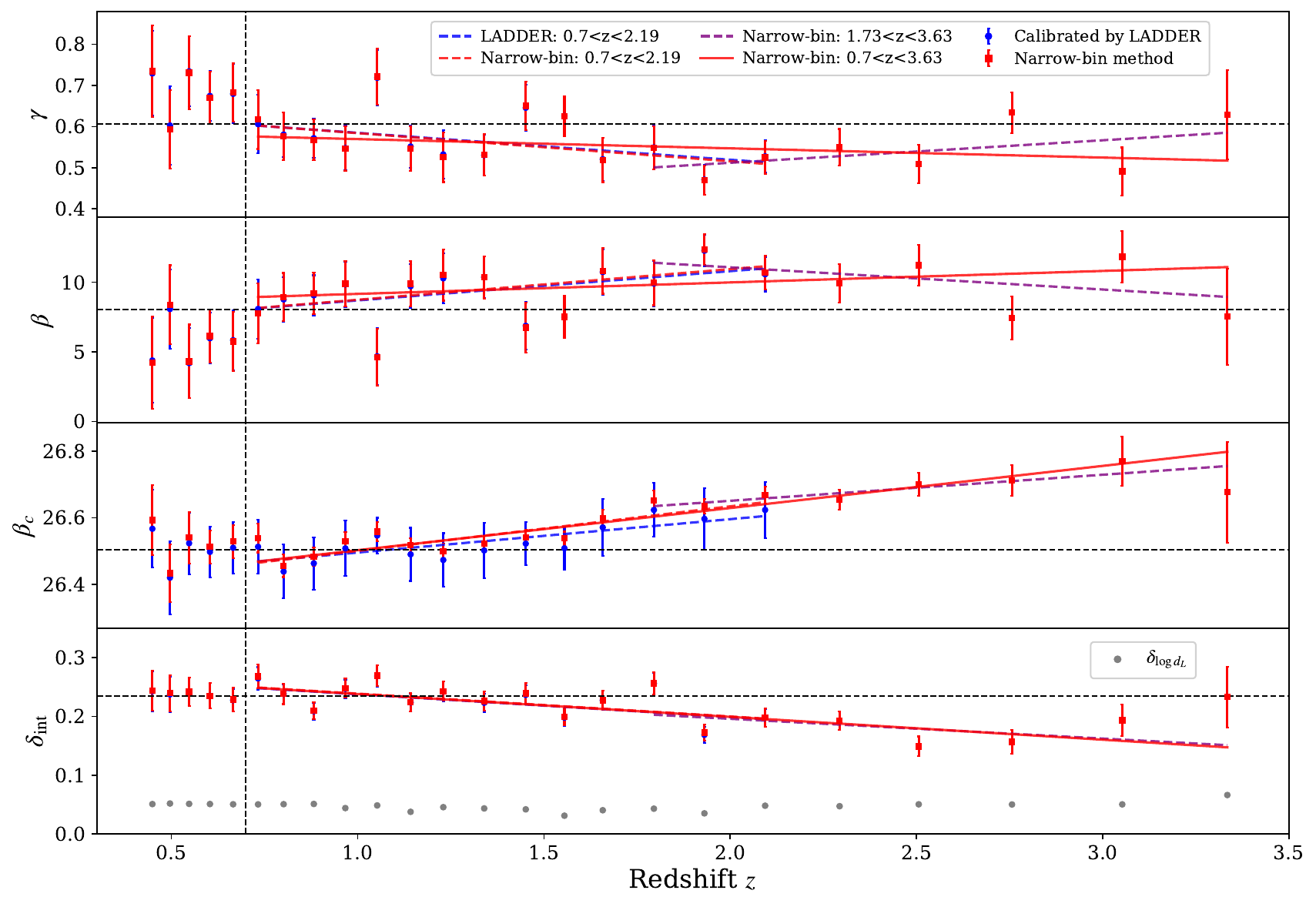}
    \caption{Comparison of scaling-relation parameters from \texttt{LADDER} (blue points) and the narrow-bin method (red points), using the bins in Table~\ref{table:3}. The $\gamma$, $\beta$, and $\delta_{\mathrm{int}}$ panels show the original fits, while $\beta_c$ shows separate centered fits. For the narrow-bin method, the luminosity intercepts and $\delta_{\log d_L}$ are calculated assuming $\Lambda$CDM ($H_0=70$, $\Omega_{\mathrm{m0}}=0.3$). Gray points in the bottom panel show $\delta_{\log d_L}$. Black horizontal dashed lines mark the posterior means for the classic relation over $0.7<z<1.70$ in Table~\ref{table:2}. All linear fits use the parameterization $p(z)=p_0+p_1z$, where $p\in \{\gamma,\beta,\beta_c,\delta_{\mathrm{int}}\}$. The red dashed, purple dashed, and red solid lines are linear fits to the narrow-bin results over $0.7<z<2.19$, $1.73<z<3.63$, and $0.7<z<3.63$, respectively. The blue dashed lines are linear fits to the \texttt{LADDER} results over $0.7<z<2.19$. \texttt{LADDER} results are shown only for bins within the SNe Ia reconstruction range, $z<2.26$.}
    \label{fig:lader_dl}
\end{figure*}

The results for both the original and centered parameterizations are listed in Table~\ref{table:3}, and the parameter comparisons are shown in Fig.~\ref{fig:lader_dl}. For the original narrow-bin fits, Table~\ref{table:3} reports the directly fitted flux intercept $\hat\beta$, whereas Fig.~\ref{fig:lader_dl} shows the luminosity intercept $\beta$ obtained through Eq.~(\ref{eq:beta_hat}). Since the centered constraints on $\gamma$ and $\delta_{\mathrm{int}}$ remain close to the original results, Fig.~\ref{fig:lader_dl} shows the original fits for $\gamma$, $\beta$, and $\delta_{\mathrm{int}}$, with the third panel displaying $\beta_c$ from separate centered fits, representing the intercept at a common reference UV luminosity.

Figure~\ref{fig:lader_dl} shows that the estimates of $\gamma$, $\beta$, $\beta_c$, and $\delta_{\mathrm{int}}$ from the two methods agree within their quoted posterior uncertainties over the common redshift range. This agreement indicates that the different treatments of distance yield compatible scaling-relation constraints over the common redshift range, strengthening our confidence in using \texttt{LADDER} on smaller data sets and applying the narrow-bin method to large high-redshift samples. Centering also reduces the posterior correlation between the slope and intercept: the median across bins of the absolute posterior Pearson correlation coefficient decreases from $0.9989$ to $0.218$ for \texttt{LADDER}, and from $0.9999$ to $0.709$ for the narrow-bin method.

Compared with the wide-bin fits, the narrow-bin results show the statistical properties of the low-redshift subsample more directly. Both fitting methods tend to yield larger $\gamma$ values in the bins at $z<0.7$, suggesting a difference in the scaling relation between this subsample and the higher-redshift sample. This finding is consistent with previous discussions of possible host-galaxy contamination in low-redshift quasars or differences between low- and high-redshift samples~\citep{Lusso:2020pdb,Lusso:2025bhy}.

The narrow-bin fits yield generally lower $\gamma$ estimates over $1.7<z<2.26$, qualitatively consistent with the redshift-dependent structure identified by the wide-bin fits over $0.7<z<2.26$. However, this decreasing trend does not persist at $z>2.26$. Visual inspection also suggests that the scatter of the bin-by-bin $\gamma$ estimates around the mean within $1.7<z<2.26$ is not noticeably larger than that within $0.7<z<1.4$. Over the broader range $0.7<z<3.63$, $\gamma$ shows no significant overall linear trend with redshift. This result for the overall slope trend is qualitatively consistent with the narrow-bin analysis of \citet{Lusso:2025bhy}.

The original intercept $\beta$ likewise shows no significant overall linear trend over the full range $0.7<z<3.63$. In contrast, the centered intercept $\beta_c$ shows an overall increasing trend. The two methods agree in the direction of this trend over their common redshift range, with larger uncertainties on $\beta_c$ for \texttt{LADDER}, which marginalizes over $H_0$ and includes reconstructed-distance uncertainties.

Given the strong posterior correlation between the original parameters, the opposite trends of $\gamma$ and $\beta$ cannot be regarded as independent evidence for evolution. Centering clarifies the interpretation of the intercept while preserving the statistical strength of the slope trends. However, the luminosity intercept inferred with the narrow-bin method still depends on the reference distance relation, so its redshift behavior must be interpreted in the context of the adopted distance calibration. This is consistent with the discussion by~\citet{Lusso:2025bhy} of the degeneracy between intercept evolution and the distance--redshift relation.

The narrow-bin results also show an overall decreasing trend in the intrinsic dispersion. A quantitative analysis of these redshift trends, together with an assessment of the impact of dispersion modeling on the scaling-relation constraints, is presented in Sec.~\ref{sec:nonlinear}.

\subsection{\label{sec:nonlinear}Redshift Trends and the Impact of Intrinsic Dispersion}

Building on the bin-by-bin comparison above, we further analyze the redshift trends of the scaling-relation parameters and examine the sensitivity of the results to intrinsic-dispersion modeling. We perform linear regressions on the bin-by-bin results from the narrow-bin method and \texttt{LADDER}. The narrow-bin method allows us to extend this analysis to higher redshifts. We consider the two subintervals $0.7<z<2.19$ and $1.73<z<3.63$, as well as the full interval $0.7<z<3.63$. The fitting results are listed in Table~\ref{table:4}, and the corresponding regression lines are shown in Fig.~\ref{fig:lader_dl}.

Under the original parameterization, the linear regression slopes for $\gamma$ and $\beta$ within $0.7<z<2.19$ differ from zero by more than $2\sigma$ for both \texttt{LADDER} and the narrow-bin method, with consistent results between the two methods. Within $1.73<z<3.63$, the narrow-bin estimates of $\gamma_1$ and $\beta_1$ differ from zero by only about $1.33\sigma$ and $1.24\sigma$, respectively. Over the full range $0.7<z<3.63$, $\gamma_1$ and $\beta_1$ differ from zero by about $1.24\sigma$ and $1.47\sigma$, respectively.

\begin{table*}[htbp]
\renewcommand\arraystretch{1.3}
\setlength{\tabcolsep}{4.7pt}
\footnotesize
\centering
\begin{tabular}{ccccccccc}
\hline
\multicolumn{9}{c}{\textbf{Original}} \\
\hline
Model & Subsamples & $\mathcal{N}$
& \multicolumn{2}{c}{$\gamma$}
& \multicolumn{2}{c}{$\beta$}
& \multicolumn{2}{c}{$\delta_{\mathrm{int}}$} \\
\cline{4-9}
& & & $\gamma_0$ & $\gamma_1$
& $\beta_0$ & $\beta_1$
& $\delta_{\mathrm{int},0}$ & $\delta_{\mathrm{int},1}$ \\
\hline
\texttt{LADDER} & $0.7<z<2.19$ & 14
& $0.650\pm0.046$ & $-0.065\pm0.031$
& $6.595\pm1.402$ & $2.100\pm0.938$
& $0.276\pm0.014$ & $-0.039\pm0.010$ \\
\hline
narrow-bin & $0.7<z<2.19$ & 14
& $0.652\pm0.047$ & $-0.068\pm0.031$
& $6.539\pm1.430$ & $2.201\pm0.947$
& $0.278\pm0.015$ & $-0.039\pm0.010$ \\
\hline
narrow-bin & $1.73<z<3.63$ & 8
& $0.403\pm0.096$ & $0.055\pm0.041$
& $14.257\pm2.984$ & $-1.593\pm1.286$
& $0.263\pm0.039$ & $-0.034\pm0.017$ \\
\hline
narrow-bin & $0.7<z<3.63$ & 19
& $0.592\pm0.034$ & $-0.022\pm0.018$
& $8.339\pm1.031$ & $0.824\pm0.559$
& $0.277\pm0.011$ & $-0.039\pm0.006$ \\
\hline\hline
\multicolumn{9}{c}{\textbf{Centered}} \\
\hline
Model & Subsamples & $\mathcal{N}$
& \multicolumn{2}{c}{$\gamma$}
& \multicolumn{2}{c}{$\beta_c$}
& \multicolumn{2}{c}{$\delta_{\mathrm{int}}$} \\
\cline{4-9}
& & & $\gamma_0$ & $\gamma_1$
& $\beta_{c,0}$ & $\beta_{c,1}$
& $\delta_{\mathrm{int},0}$ & $\delta_{\mathrm{int},1}$ \\
\hline
\texttt{LADDER} & $0.7<z<2.19$ & 14
& $0.657\pm0.046$ & $-0.070\pm0.031$
& $26.395\pm0.071$ & $0.100\pm0.052$
& $0.277\pm0.015$ & $-0.039\pm0.010$ \\
\hline
narrow-bin & $0.7<z<2.19$ & 14
& $0.652\pm0.047$ & $-0.068\pm0.031$
& $26.367\pm0.026$ & $0.134\pm0.018$
& $0.278\pm0.015$ & $-0.039\pm0.010$ \\
\hline
narrow-bin & $1.73<z<3.63$ & 8
& $0.394\pm0.098$ & $0.059\pm0.042$
& $26.494\pm0.080$ & $0.078\pm0.037$
& $0.262\pm0.039$ & $-0.033\pm0.017$ \\
\hline
narrow-bin & $0.7<z<3.63$ & 19
& $0.591\pm0.034$ & $-0.021\pm0.018$
& $26.376\pm0.020$ & $0.126\pm0.013$
& $0.277\pm0.011$ & $-0.039\pm0.006$ \\
\hline
\end{tabular}
\caption{Linear regression results for the bin-by-bin scaling-relation parameters, using $p(z)=p_0+p_1z$. Original and Centered use the extrapolated intercept $\beta$ and the normalization $\beta_c$ at the fixed sample mean of $\log L_{\mathrm{UV}}$, respectively, as in Table~\ref{table:3}. The other parameters are $\gamma$ and $\delta_{\mathrm{int}}$. Here $\mathcal{N}$ is the number of bins used in each regression. Regressions use inverse-variance weights based on the bin posterior standard deviations, treating the bins as independent. The quoted regression errors are not rescaled by the reduced chi-squared.}
\label{table:4}
\end{table*}

The fitted linear trends of both $\gamma$ and $\beta$ reverse sign between $0.7<z<2.19$ and $1.73<z<3.63$, suggesting possible non-monotonic redshift behavior. The evidence for a trend is stronger within $0.7<z<2.19$ and weaker at higher redshifts, but remains limited in both subintervals. These opposite trends may partly cancel in the full-range linear fit, consistent with the absence of a significant overall linear trend over $0.7<z<3.63$.

For the centered intercept $\beta_c$, the narrow-bin method yields a significant positive linear trend over $0.7<z<3.63$ under the adopted reference $\Lambda$CDM distance relation and regression assumptions.

In addition to the trends in the scaling-relation parameters, the narrow-bin results show a decreasing trend in the intrinsic dispersion, motivating a test of how dispersion modeling affects the scaling-relation constraints. We refit the classic and redshift-corrected relations to the $0.7<z<2.26$ sample, introducing two separate parameters, $\delta_{\mathrm{int,low}}$ and $\delta_{\mathrm{int,high}}$, with a division at $z=1.6$. Comparing Tables~\ref{table:2} and~\ref{table:5}, the shifts in the scaling-relation parameters are small relative to their posterior widths. The log Bayes factors favoring the redshift-corrected relation are approximately $16.40$ and $18.32$ under the shared- and split-dispersion assumptions, respectively. Both settings strongly favor the redshift-corrected relation, indicating that the original model preference is robust to the dispersion parameterization tested here.

\begin{table*}[htbp]
\renewcommand{\arraystretch}{1.6}
\setlength{\tabcolsep}{4pt}
\footnotesize
\centering
\caption{Fitting results for $0.7<z<2.26$ using the same data
and fitting settings as in Table~\ref{table:2}, but with separate
intrinsic dispersions $\delta_{\mathrm{int,low}}$ and
$\delta_{\mathrm{int,high}}$ for $z<1.6$ and $z\geq1.6$,
respectively.}
\label{table:5}
\begin{tabular}{ccccccccc}
\hline
Subsample & Relation
& $\gamma$ ($\gamma_0$) & $\beta$ ($\beta_0$)
& $\gamma_1$ & $\beta_1$
& $\delta_{\mathrm{int,low}}$
& $\delta_{\mathrm{int,high}}$
& $\ln\mathcal{Z}$ \\
\hline\hline
\multirow{2}{*}{$0.7<z<2.26$}
& classic relation
& $0.613\pm0.011$ & $7.876\pm0.346$
& $\ldots$ & $\ldots$
& $0.234\pm0.005$ & $0.227\pm0.008$
& $0.670$ \\
\cline{2-9}
& $z$-corrected
& $0.657\pm0.065$ & $6.234\pm1.969$
& $-0.043\pm0.026$ & $1.423\pm0.803$
& $0.233\pm0.005$ & $0.214\pm0.008$
& $18.992$ \\
\hline
\end{tabular}
\end{table*}

In our parallel study~\citep{2026arXiv260627173G}, we also found that the redshift trend of the intrinsic dispersion remains stable under the scaling-relation parameterizations tested. Together, these results indicate that, within the models considered, the redshift behavior of the scaling relation and that of the intrinsic dispersion are robust to changes in each other's parameterization. For the $1.6<z<2.6$ sample, that study assumed a flat $\Lambda$CDM cosmology, fixed the scaling-relation calibration parameters, and adopted the SH0ES prior on $H_0$, obtaining posterior medians of $\Omega_{\mathrm m0}=0.298$ and $0.323$ for the constant and redshift-dependent dispersion models, respectively. This shift of approximately $0.025$ indicates that dispersion modeling should be included in assessments of systematic uncertainties. However, within this controlled setup, its impact is insufficient to explain the larger parameter differences between some quasar analyses and other cosmological probes.

Taken together, our results support redshift-dependent statistical structure in both the scaling relation and its intrinsic dispersion. In the narrow-bin analysis, $\gamma$ shows no significant overall linear trend over $0.7<z<3.63$, while the local trends suggest possible non-monotonic behavior with limited statistical evidence. Given the impact of the scaling-relation parameters on distance and cosmological parameter estimates~\citep{Hu:2022udt}, the origin of these features deserves further investigation. For the narrow-bin method, the inferred redshift behavior of the luminosity intercepts $\beta$ and $\beta_c$ also depends on the adopted cosmology through the intercept--distance degeneracy. These findings do not conflict with the wide-bin preference for the redshift-corrected relation within $0.7<z<2.26$: the absence of a significant overall linear trend in $\gamma$ does not establish the redshift invariance of the full scaling relation, and the two analyses address different statistical questions over different redshift ranges. The redshift behavior of the scaling relation and its intrinsic dispersion may share a common physical or observational origin, including sample selection effects. Studying these features jointly can help improve quasar calibration and error modeling.

\section{\label{sec:summary}Conclusions}

In this work, we use the deep learning algorithm \texttt{LADDER} to reconstruct the Hubble diagram from the Pantheon+ SNe Ia sample and investigate the redshift dependence of the empirical quasar X-ray/UV luminosity relation. We assess the consistency of the results through comparisons with GP regression and the narrow-bin method.

Our main conclusions are as follows:
\begin{enumerate}
\item[(i)] The wide-bin results from \texttt{LADDER} are consistent with those from GP regression, and its bin-by-bin parameter constraints agree with those from the narrow-bin method. Over the common redshift range, the directions of the fitted narrow-bin trends are also qualitatively consistent with the wide-bin results. These comparisons support the robustness of our results to the choice of analysis method.

\item[(ii)] Both the wide-bin and narrow-bin analyses show that including the $z<0.7$ sample affects the scaling-relation fits, providing further support for previously reported concerns about low-redshift quasars~\citep{Lusso:2020pdb,Lusso:2025bhy}. Possible causes include host-galaxy UV contamination or differences between the samples. These data therefore require further selection or an assessment of systematic uncertainties before use in cosmological analyses.

\item[(iii)] Within $0.7<z<2.26$, the Bayesian evidence strongly favors the redshift-corrected relation, and this model preference remains unchanged under the split-dispersion prescription tested here. In the narrow-bin analysis, $\gamma$ shows decreasing and increasing trends over $0.7<z<2.19$ and $1.73<z<3.63$, respectively, with the corresponding regression slopes differing from zero by about $2.19\sigma$ and $1.33\sigma$. No significant overall linear trend is found over the full range $0.7<z<3.63$. These results suggest possible non-monotonic redshift behavior, although the statistical evidence for the local trends remains limited.

\item[(iv)] Centering reduces the posterior correlation between the slope and intercept while leaving the constraints on $\gamma$ nearly unchanged. The centered intercept $\beta_c$ shows an overall increasing trend. The opposite trends of $\gamma$ and $\beta$ in the original parameterization cannot be regarded as independent evidence for evolution. Moreover, the luminosity intercept inferred with the narrow-bin method depends on the reference distance relation, so its redshift behavior must be interpreted in the context of the adopted distance calibration. This is consistent with the discussion by \citet{Lusso:2025bhy} of the degeneracy between intercept evolution and the distance--redshift relation.
\end{enumerate}

These redshift-dependent features may be affected by sample selection effects and do not, by themselves, rule out the use of quasars as standardizable candles. Identifying their origin and accounting for them in calibration and error modeling can help make fuller use of the large quasar sample and its broad redshift coverage, improving the reliability of quasar cosmology.

\begin{acknowledgments}
  This work has been supported by the National Key Research and Development Program of China (No. 2022YFA1602903) and the National Natural Science Foundation of China (Nos.~12473002, 12075042 and 12475047). We used ChatGPT (6 Astra, OpenAI) to assist with code review, writing plotting code, consistency checks, discussions of statistical and physical interpretations, and the translation and language editing of selected passages. The authors verified the relevant outputs and take full responsibility for this work.
\end{acknowledgments}

\bibliography{apssamp}
\end{document}